# A New Imaginary Term in the 2$^{nd}$ Order Nonlinear Susceptibility from Charged Interfaces

Emily Ma,[#] Paul E. Ohno,[$] Jeongmin Kim,[%] Yangdongling Dawning Liu,[#] Emilie H. Lozier,[#]

Thomas F. Miller III,[%] Hong-Fei Wang,[†@] and Franz M. Geiger[#]*

[#]Department of Chemistry, Northwestern University, 2145 Sheridan Road, Evanston, IL 60660,

USA

[$]Harvard University Center of the Environment, Harvard University, Cambridge, MA 02138,

USA

[%]Division of Chemistry and Chemical Engineering, California Institute of Technology,

Pasadena, CA 91125, USA

[†]Department of Chemistry and Shanghai Key Laboratory of Molecular Catalysis and Innovative

Materials, Fudan University, 220 Handan Road, Shanghai 200433, China

[@]School of Sciences, Westlake University, Shilongshan Road No.18, Cloud Town, Xihu District,

Hangzhou, Zhejiang 310024, China

**Abstract.** Non-resonant second harmonic generation phase and amplitude measurements obtained from the silica:water interface at varying pH and 0.5 M ionic strength point to the existence of a nonlinear susceptibility term, which we call $\chi_X^{(3)}$, that is associated with a 90° phase shift. Including this contribution in a model for the total effective second-order nonlinear susceptibility produces reasonable point estimates for interfacial potentials and second-order nonlinear susceptibilities when $\chi_X^{(3)} \approx 1.5 \times \chi_{water}^{(3)}$. A model without this term and containing only traditional $\chi^{(2)}$ and $\chi^{(3)}$ terms cannot recapitulate the experimental data. The new model also provides a demonstrated utility for distinguishing apparent differences in the second-order nonlinear susceptibility when the electrolyte is NaCl vs MgSO$_4$, pointing to the possibility of using HD-SHG to investigate ion-specificity in interfacial processes.

*Corresponding author: geigerf@chem.northwestern.edu

We recently reported a heterodyne-detected second harmonic generation (HD-SHG) spectrometer for simultaneously determining the amplitude and phase of nonlinear optical signals generated at solid:aqueous interfaces.[1] The model underlying the data analysis assumed two contributions, one in the form of the second-order nonlinear susceptibility, $\chi^{(2)}$, from the interface, and one from the product of the surface potential and an effective third-order nonlinear susceptibility response of the diffuse layer, $\Phi(0) \times \chi^{(3)}_{EDL}$. We now present SHG amplitude and phase measurements carried out at the fused silica:water interface that indicate the necessity of the presence of an additional contribution to quantitatively describe the total nonlinear optical response. This additional contribution varies with the surface potential and shifted in phase by 90° relative to the second-order nonlinear susceptibility. The inclusion of this new term as the sole adjustable parameter yields a model that produces point estimates for the second-order nonlinear susceptibility, $\chi^{(2)}$, and the interfacial potential, $\Phi(0)$. These point estimates are in reasonable agreement with those reported by independent methods under similar conditions of varying pH and an ionic strength of 0.5 M NaCl, as well as varying ionic strength at constant pH, when the new term is similar in value to the third-order nonlinear susceptibility of water.

In the experiments, we record the SHG amplitude and phase from fused silica hemispheres in contact with 0.5 M aqueous NaCl solutions held between pH 1 and 12 adjusted using HCl and NaOH. We also vary the ionic strength at a constant pH of 5.8. We use 40 mW of input power from a Pharos amplifier laser system operating at 200 kHz and 1030 nm, with 190 femtosecond pulses. The data are obtained in replicates using several fused silica hemispheres. We reference the number of SHG photons detected from the silica:water interface at zero added NaCl and pH 5.8 (ionic strength from conductivity measurements = 2 $\mu$M) against that obtained from the fused silica:z-cut $\alpha$-quartz interface and account for Fresnel coefficients using our previously published approach.[2] Interference fringes are recorded using a second piece of z-cut $\alpha$-quartz as a local oscillator on a 100 mm long translational stage, with each measurement taking about four minutes, as described previously.[1-3] Each mm distance equals 3.158° in phase space, allowing us to map out close to 90 percent of a given fringe.

Fig. 1a shows that the SHG fringes exhibit smaller offsets and amplitudes as the pH is lowered, matching the first report of homodyne-detected measurements of SHG intensities by Eisenthal and co-workers.[4] Fig. 1b shows the nonlinear optical response in our spectrometer depends quadratically on in input power, both in homodyne and heterodyne detection. Computing the square of the heterodyne-derived SHG amplitudes results in what would be the homodyne-detected SHG intensity, which, as shown in Supporting Information Fig. S1, recapitulates the earlier reported homodyne-detected SHG measurements under identical conditions in the aqueous phase.[4] Fig. 1c shows the results for pH 1 to pH 12, with the fringe extrema falling on different parts of the 100 mm time delay stage due to day-today differences in absolute phasing. Fig. 1d shows the SHG phases and amplitudes we obtained for the pH 1 to pH 12 range studied and indicates that both depend strongly on pH at this high ionic strength. We set the SHG phase at pH 2.5 and 0.5 M NaCl to zero, given that this pH corresponds to the point of zero charge of fused silica[5-8] and thus should most closely approximate a purely $\chi^{(2)}$ response. Qualitatively similar results are obtained at an ionic strength of 50 mM (Supporting Information Fig. S2, note that a narrower pH range was studied so as to maintain constant ionic strength). Fig. 1e shows replicate measurements of SHG amplitude and phase at pH 5.8 and 0.5 M NaCl taken from separate trials taken on separate days which show a standard deviation of ± 4 percent for the amplitude and ± 1° for the phase measurements.

Under the non-resonant conditions of our experiment ($\lambda_\omega$=1030 nm, $\lambda_{2\omega}$=515 nm), the second- and third-order nonlinear susceptibilities are purely real,[9] as is the surface potential.[10] The SHG response is given by the sum of the second- and third-order contribution according to

$$\chi_{tot}^{(2)} = \chi^{(2)} - \chi_{EDL}^{(3)} \Phi(0) \cos\left(\varphi_{DC,EDL}\right) e^{i\varphi_{DC,EDL}} \qquad (1).$$

Here, the phase angle associated with the DC field emanating from the interface into the electrical double layer (EDL) is given by $\varphi_{DC,EDL} = atan(\Delta k_z \lambda_D)$,[1, 11-20] where the wave vector mismatch, $\Delta k_z$, is 1.1 x $10^7$ m$^{-1}$ in our experimental setup[1-2] and $\lambda_D$ is the Debye screening length in the aqueous phase computed from Debye Hückel theory. As the Debye screening length is 4 Å at 0.5 M ionic strength, the DC phase angle at that condition is 0.27°. Taking $\chi_{EDL}^{(3)} = \chi_{water}^{(3)}$=1x$10^{-21}$ m$^2$V$^{-2}$, obtained either from heterodyned non-resonant SHG experiments[21] or from an estimate computed using quantum mechanical inputs,[13] the total second-order response is essentially purely real, i.e. $\chi_{tot}^{(2)} = \chi^{(2)} - \chi_{water}^{(3)}\Phi(0)$. This expression matches that of the first report on this topic,[4] though we have altered the sign convention to match that used recently by other researchers in the field.[12, 22] The measured SHG response, which is given by its amplitude and the phase in the form of $E_{sig}e^{i\varphi_{sig}}$, should then be purely real (phase of 0° or 180°) at these short Debye screening lengths, no matter the values of $\chi^{(2)}$, $\chi_{water}^{(3)}$, or $\Phi(0)$. However, Fig. 1d shows that $\varphi_{sig}$ changes when going from low to high pH, for which $\Phi(0)$ is well-documented to vary for fused silica.[5-7]

This variation of the measured SHG phase with pH, and thus $\Phi(0)$, indicates the presence of an additional complex-valued contribution to the $\chi_{tot}^{(2)}$ response.

To quantify this newly identified term, we added the most general form of a complex-valued contribution to $\chi_{tot}^{(2)}$ in eqn. 1, as expressed below:

$$E_{sig} \times e^{i\varphi_{sig}} = \chi_{tot}^{(2)} = \chi^{(2)} - \Phi(0)\chi_{water}^{(3)} \cos(\varphi_{DC,EDL}) e^{i\varphi_{DC,EDL}} - \chi_X^{(n)} \quad (2)$$

Collection of the imaginary parts yields

$$\chi_X^{(n)} = -E_{sig} \times \sin\varphi_{sig} - \Phi(0)\chi_{EDL}^{(3)} \cos(\varphi_{DC,EDL}) \sin(\varphi_{DC,EDL}) \qquad (3)$$

Fig. 1f shows that $\chi_X^{(n)}$ is linearly correlated with surface potentials, $\Phi_{imped}$, reported by Diot et al.[23] from impedance measurements performed as a function of pH at 1 M NaCl, close to the 0.5 M ionic strength employed in our present study. The impedance study reported the point of zero charge to be around 2.5, as it is generally accepted to be for silica,[5-8] and a positive potential of *ca.* +10 mV at pH 1. The potential at pH 5.8 was found to be *ca.* -80 mV, while it reached *ca.* -180 mV at pH 8 and *ca.* -310 mV at pH 11. Similar values were reported for acidic pH values using silica-terminated field effect devices reported by Bousse et al.[8] for 1M and 0.1 M NaNO$_3$. Discounting the caveat that the nitrate anion has been shown to be surface active for silica by resonantly enhanced SHG measurements,[24] we can also compare our results to the potentials reported in that work (+15 mV at pH 1, between 0 and -8 mV for pH 2 to pH 3, -25 mV at pH 4, and -40 mV at pH 5). Taken together, the linear correlation evident in Fig. 1f can be summarized by $\chi_X^{(n)} = \chi_X^{(m)}\Phi_{imped}$. The units of the newly identified contribution $\chi_X^{(m)}$ must match that of $\chi_{tot}^{(2)}$ (m$^2$V$^{-1}$), which is satisfied if the units of $\chi_X^{(m)}$ are (m$^2$V$^{-2}$), indicative of a third-order susceptibility, i.e. $\chi_X^{(m)} = \chi_X^{(3)}$. Linear least squares fitting of the data shown in Fig. 1f yields $\chi_X^{(3)} = 1.55 \times 10^{-21} \pm 7 \times 10^{-23}\ m^2V^{-2}$.

While the existence of this additional term is established, its origin needs to be discussed in the following.

In earlier work,[16] we had stated "In addition, the surface potential can not only induce bulk $\chi^{(3)}$ responses from the water side, but also from the fused silica or the α-quartz side.[25] These issues warrant further investigation in the future." We were motivated by electric field induced second harmonic generation from p-doped silicon across a 9 nm thin layer of $SiO_2$,[26] and, furthermore, the considerably larger SHG intensity difference for high vs low pH conditions that is obtained from fused silica:water[4, 27-28] (also Supporting Information Fig. S1) vs air:water interfaces that contain monolayers of surfactants with ionizable headgroups.[29-30] We write, in analogy to the derivation for the dc contribution on the aqueous side in eqn. (1),[1-2, 11-12, 15, 31] the following expression for the dc contribution on the solid silica side:

$$\chi^{(2)}_{dc,SiO_2} = \chi^{(3)}_{SiO_2} \int_{-\infty}^{0} E_{DC}(z) e^{i\Delta k_z z} dz \qquad (4)$$

Here, $E_{DC}(z) = -\frac{\Phi(z)}{dz} (z \leq 0)$, with $\Phi(z)$ decaying linearly from its value at the aqueous:solid interface, $\Phi(0)$, to zero at the edge of the hemisphere, taken to be the electrical ground. In other words, $\Phi(z) = $ a(z + b) $(for -b \leq z \leq 0)$ and $\Phi(z) = 0$ $(for\ z \leq -b)$, where b = 1.27 cm, the radius of our fused silica hemisphere, and $a = \frac{\Phi(0)}{b}$. The solution to eqn 4 is

$$\chi^{(2)}_{dc,SiO_2} = -\chi^{(3)}_{SiO_2} \frac{\Phi(0)}{b} \int_{-b}^{0} e^{i\Delta k_z z} dz = -i\chi^{(3)}_{SiO_2} \frac{\Phi(0)}{b\Delta k_z} \left( e^{-i\Delta k_z b} - 1 \right) \quad (5).$$

Eqn. 1 now becomes (please see Supporting Information Note S1)

$$\chi^{(2)}_{tot} = \chi^{(2)} - \Phi(0) \left( \chi^{(3)}_{water} \cos(\varphi_{DC,EDL}) e^{i\varphi_{DC,EDL}} - \chi^{(3)}_{SiO_2} \frac{\sin(\varphi_{DC,SiO_2})}{\varphi_{DC,SiO_2}} e^{-i\varphi_{DC,SiO_2}} \right) \quad (6),$$

where $\varphi_{DC,SiO_2} = \frac{\Delta k_z b}{2} = 7 \times 10^4$. Recent experimental work reports $\chi^{(3)}_{SiO_2}$=2 x $10^{-22}$ $m^2V^{-2}$.[32] Accordingly, with $\chi^{(3)}_{water}$=1x10$^{-21}$ $m^2V^{-2}$, and $\varphi_{DC,EDL}$=0.27° at 0.5M [NaCl], we find that the third-order contribution from water dominates that of the silica by several orders of magnitude, essentially returning to eqn. 1.

Fused silica contains impurities, which we considered using one-dimensional finite element calculations (0.01 nm steps, see Supporting Information Note S2) of a solid having a relative permittivity, ε, of 3.8,[33-34] and 100 ppb charged impurities[35] with vacuum on one side and a 0.1M aqueous salt solution on the other side. The surface charge density is set to -0.015 C m$^{-2}$, and the relative permittivity of the 0.1 M salt solution is taken to be 78.[34] Averaged over 200 nm of solid, and 10$^6$ simulations, the model shows the electric field is -100 kV m$^{-1}$ ± 2 MV m$^{-1}$, *i.e.* statistically indistinguishable from zero (Fig. 2A, B). Nevertheless, using the point estimate of -100 kV m$^{-1}$ between z=0 nm, the laser focus position, to z=-b=-0.0127 m, the hemisphere's edge, we obtain $\chi_{dc,SiO_2}^{(2)} = 8 \times 10^{-25} \times e^{-145°i} \, m^2 V^{-2}$, which is again much smaller than the $\chi_{water}^{(3)}$ contribution. (Note that we can also integrate $E_{DC}(z)$ obtained from the finite element calculations over 200 nm, ~2 x the coherence length in our experiments, which yields -0.013 V, three times larger than the result obtained using the field point estimate of -100 kV m$^{-1}$, arriving at the same conclusion). Our two analyses indicate that the bulk silica contribution is not enough to explain the pH-dependent SHG phase observed at short Debye lengths (0.5 M [NaCl], Fig. 1d), at least within the electric dipole approximation.

We also considered the impact on the third-order contribution due to the finite roughness of the surface of the fused silica hemispheres. Atomic force microscopy shows the surface to have an rms roughness of 0.4 nm, with +/- 1 nm variation in surface height (Fig. 2C). We therefore assigned the aqueous:solid interface a width, ω, of 1 nm, as opposed to being atomically flat. We also assign this first nm of liquid a relative permittivity of 3.[36-37] We treat this 1 nm wide region with a third-order susceptibility taken to be the geometric mean of $\chi_{water}^{(3)}$ and $\chi_{SiO_2}^{(3)}$, i.e. $\chi_{\varpi}^{(3)}$=6 x $10^{-22}$ $m^2V^{-2}$. The second-order dc-contribution is then

$$\chi_{dc,\varpi}^{(2)} = \chi_{\varpi}^{(3)} \int_{\varpi=-1\,nm}^{0} E_{DC}(z) e^{i\Delta k_z z} dz \qquad (7)$$

where $E_{DC}(z)$ can be approximated, again guided by finite element calculations (Fig. 2D), to be constant at $E_{DC}(z) = -\frac{\Phi(0)}{1\times10^{-9}} Vm^{-1}$ over the 1-nm wide rough interfacial region. The second-order dc contribution to the second order nonlinear susceptibility then becomes $\chi_{dc,\varpi}^{(2)} = -\Phi(0) \times \frac{6\times10^{-22} m^2V^{-2}}{1\times10^{-9}m} \int_{\varpi=-1\,nm}^{0} e^{i\Delta k_z z} dz = -\Phi(0) \times 6 \times 10^{-22} \times e^{-0.3°i}\ m^2V^{-1}$ , having, again, an imaginary part that is insufficient to explain the observed pH-dependent SHG phase shift at high ionic strength.

We also considered multipolar terms, which, according to atomistic simulations, can be important for interfacial potentials of aqueous interfaces.[38-40] These terms also act as important contributors to nonlinear optical responses[41] under some conditions, given that they are bulk allowed.[42-45] Our SHG experiment uses a single-beam input geometry, where quadrupolar responses to the second-order nonlinear susceptibility should not be observable.[46] Yet, this rule may break down somewhat given the surface roughness of our fused silica hemispheres (*vide supra*) and, perhaps, non-ideal focusing and wavefronts of the fundamental at the interface. We therefore tested whether adding an electrical quadrupolar term, $\chi_q^{(2)}$ , that is out of phase by $\pm90°$=$\pm i = e^{\pm i\frac{\pi}{2}}$ from $\chi^{(2)}$ in eqn. 1, and is furthermore independent of the surface potential, could be important, according to $\chi_{tot}^{(2)} = \chi^{(2)} \pm i\chi_q^{(2)} - \chi_{EDL}^{(3)}\Phi(0) \cos(\varphi_{DC,EDL}) e^{i\varphi_{DC,EDL}}$ . The $90°$ phase shift originates from integration of this bulk-allowed response over the coherence length of our experimental setup, *i.e.* $\Delta k_z \times \int_0^\infty e^{\pm i\Delta k_z z} dz = \pm i$, to account for the fact that they are bulk allowed. Setting $\chi_q^{(2)} \approx \chi^{(2)}$,[41] we find a sizeable imaginary contribution to $\chi_{tot}^{(2)}$ at high ionic strength that is, however, invariant with surface potential, as posited in this model. As shown in Supporting Information Note S3, Fig. S5, and Fig. S6, the resulting $\chi^{(2)}$ and $\Phi(0)$ point estimates are non-physical. This outcome is not surprising, given the strong linear correlation of $\chi_X^{(n)}$ with surface potential we determined from regressing our experimental SHG amplitude and phase data against the surface potentials determined from impedance measurements (Fig. 1f).

As pointed out earlier, the units of the quadrupolar second-order nonlinear susceptibility ($m^2V^{-1}$) prevent its coupling to the surface potential as the multiplication of the units of any new term $\chi_X^{(n)}$ and $\Phi(0)$ [V] must equal those of the total second-order nonlinear susceptibility, $\chi_{tot}^{(2)}$ ($m^2V^{-1}$). Thus, the units of the new term must be ($m^2V^{-2}$). Multipolar terms in the third-order nonlinear susceptibility have such units.[47] At the high ionic strengths employed in our experiments, the charged interface can essentially be viewed as a capacitor with a field that acts over the Debye length, which is <1 nm at 0.5M [salt].[5-7] Multipolar third-order contributions could be considerable in the presence of such fields. Just like in their second-order counterparts, quadrupolar third-order contributions should be associated with a 90° phase shift relative to the ones from electric dipoles and be bulk-allowed.

If one were to introduce an additional third-order contribution, $\chi_X^{(3)}$, that is of bulk origin, associated with a 90° phase shift, and coupled to the surface potential, one may propose the total second-order response from the charged interface, again at 0.5 M salt, to take the following form:

$$\chi_{tot}^{(2)} = \chi^{(2)} - \Phi(0)\left(\chi_{water}^{(3)} \cos(\varphi_{DC,EDL}) e^{i\varphi_{DC,EDL}} \pm i\chi_X^{(3)}\right) \quad (8)$$

This expression is essentially that discussed in the context of eqn. 2. Taking the magnitude of $\chi_X^{(3)}$ to be $1.5 \times \chi_{water}^{(3)}$, one finds that $\chi_{tot}^{(2)} = \chi^{(2)} - \Phi(0)\left(1 \times 10^{-21} \times \cos(0.27°) \times\right.$

$$e^{0.27^\circ i} \ m^2 V^{-2} \pm 1.5 \times 10^{-21} \times e^{90^\circ i} \ m^2 V^{-2}) = \chi^{(2)} - \Phi(0)\big(1.8 \times 10^{-21} \times e^{\pm 56^\circ i} \ m^2 V^{-2}\big) \quad .$$

This form allows for a sizeable complex valued SHG response, *i.e.*, a nonzero SHG phase, under conditions where the Debye length is compressed to the point where the $\cos(\varphi_{DC,EDL}) e^{i\varphi_{DC,EDL}}$ product is purely real.

      To test whether eqn. 8 yields reasonable estimates of surface potential and second-order nonlinear susceptibility, the structural and electrostatic parameters of interest to us, we calibrated the SHG signal intensity measured from the silica:water interface in our spectrometer against that of fused silica bonded *via* an index-matched adhesive (Nordland 146H) to a vertically aligned piece of z-cut α-quartz (no water present). z-Cut α-quartz is an IEEE standard having an optical phase of zero or 180°, depending on crystal orientation, and a known $\chi^{(2)}_{Bulk,Q}$ of $8 \times 10^{-13}$ m V$^{-1}$,[48-49] as described in our previous work.[2] After determining $E_{sig,quartz}$ from $\sqrt{I_{sig,quartz}}$, and repeating the measurement with water at pH=5.8 and 2 µM ionic strength in place of the quartz, we find the calibration and referencing ratio

$$\frac{C}{R} = \frac{1}{E_{sig,quartz}} \frac{F_{quartz}}{F_{sample}} \left| \chi^{(2)}_{eff,quartz} \right| \tag{9},$$

where $\left| \chi^{(2)}_{eff,quartz} \right| = \frac{8 \times 10^{-13} \ mV^{-1}}{\Delta k_z} = 4.5 \times 10^{-20}$ m$^2$V$^{-1}$ and $F_{quartz}$ and $F_{sample}$ are the macroscopic Fresnel coefficients (0.5 and 2.58, resp., in our experimental setup).[2] To quantify the C/R ratio, we performed seven replicate measurements of the SHG intensity from individually assembled silica:α-quartz interfaces, which yielded 42 x $10^3$ ± 1 x $10^3$cps (5 mW laser power at the sample). Seven subsequent replicates from individually assembled silica:ultrapure water interfaces (air-equilibrated over night, 20 µS cm$^{-1}$ conductivity, pH 5.8,) yielded 71 ± 1 cps. The ratio $\frac{1}{E_{sig,quartz}}$ is then 24, and the total C/R ratio is $3.6 \times 10^{-22} m^2 V^{-1}$ in our spectrometer, with an uncertainty of 5 percent (one σ, please see Supporting Information Fig. S9). We believe the main contributors to this uncertainty are minute changes in laser spot position and focus at the interface that stem from our inability to place the fused silica hemisphere onto the sample cell in the exact same position prior to each experiment.

The properties we seek, $\chi^{(2)}$ and $\Phi(0)$, are obtained as follows:[50]

$$\Phi(0) = -\frac{C}{R} \times \frac{E_{sig,sample} \sin(\varphi_{sig})}{\chi_{water}^{(3)} \cos(\varphi_{DC,EDL}) \sin(\varphi_{DC,EDL}) + \chi_X^{(3)}} \qquad (10a), \text{ and}$$

$$\chi^{(2)} = \frac{C}{R} \times \left(E_{sig,sample}\right) \cos(\varphi_{sig}) + \Phi(0) \chi_{water}^{(3)} \cos^2(\varphi_{DC,EDL}) \quad (10b).$$

At 0.5 M salt, $\varphi_{DC,EDL} = \arctan(\Delta k_z \, \lambda_D) = 0.27°$, irrespective of pH or surface chemistry. We use the recently published value[13, 21] of 1 x $10^{-21}$ m$^2$V$^{-2}$ for $\chi_{water}^{(3)}$. There are no other adjustable parameters or inputs except for the choice of the magnitude of $\chi_X^{(3)}$. The uncertainties on the $\chi^{(2)}$ and $\Phi(0)$ point estimates are taken to be dominated by the 5 percent uncertainty of the $\frac{C}{R}$ ratio.

Fig. 3a shows the $\chi^{(2)}$ and $\Phi(0)$ point estimates obtained using the measured SHG amplitude and phase from Fig. 1 in eqn. 8 with $\chi_X^{(3)}$ taken to be 1.5 times that of $\chi_{water}^{(3)}$. The corresponding Argand diagrams depicting the relevant elements of eqn. 8 in the complex plane are shown in Fig. 3b. The $\chi^{(2)}$ values obtained in this fashion are within a factor of 2 of those published for other charged aqueous interfaces, such as those reported recently by the Roke group for silica suspensions,[51] albeit for much lower ionic strengths. Point estimates of $\chi^{(2)}$ reported in that work were reported for salt concentrations of 10 mM and pH 10 to be around 2 x $10^{-22}$ m$^2$V$^{-2}$, while negative point estimates were reported for ionic strengths of 0.1 mM and below at pH 10 and pH 5.7. The pH-dependent surface potential estimates are not appropriate for comparison here, since the highest ionic strength in that work was 10 mM, whereas it is 0.5 M here. Yet, as mentioned above, Diot et al.[23] and Bousse et al.[8] obtained pH titrations using impedance measurements at 1.0 M NaCl, which we consider comparable to our experimental conditions. Fig. 3a shows that the surface potentials determined from the impedance measurements are well recapitulated here when setting $\chi_X^{(3)} = 1.5 \times \chi_{water}^{(3)}$, which leads to $\Phi(0)$=+20 mV at pH 1, $\Phi(0)$=-75 mV at pH 5.8, and $\Phi(0)$=-330 mV at pH 11. This result is consistent with the linear correlation found from Fig. 1f.

It is reassuring that the surface potentials derived by HD-SHG are larger in magnitude than the $\zeta$-potentials obtained at 0.1M and 0.01 M salt between pH 2 and pH 8,[52] which we expect given that the $\zeta$-plane is at some distance away from the zero-plane in the aqueous phase. Yet, we note the caveat that $\zeta$-potentials are not sensitive to the dipole potential, as atomistic simulations by the Netz group have shown.[53] Compared to the original homodyne-detected SHG data reported by Ong, Zhao, and Eisenthal,[4] recorded at the same ionic strength of 0.5 M salt we use here, the HD-SHG-derived $\Phi(0)$ point estimates reported here are generally three times larger. Available X-ray spectroscopic surface potentials for silica colloids suspended in a liquid jet were limited to 50 mM,[54] too low for comparison to the 0.5M results presented here. Yet, it is encouraging to find that the 50 mM pH titration shown in Supporting Information Fig. S2 yields reasonable agreement with the 50 mM titration by XPS reported in that work, using, again, $\chi_X^{(3)} = 1.5 \times \chi_{water}^{(3)}$.

To further test the utility of HD-SHG surface potential measurements at lower ionic strengths, we recorded the SHG amplitude and phase for a fused silica hemisphere in contact with aqueous solutions held at a constant pH of 5.8 as a function of salt concentration (Fig. 4a). Unlike in our previous work,[1] the silica surface at each salt concentration had not been previously exposed to 100 mM NaCl. While the data are qualitatively similar, there are quantitative differences, as expected due to previously reported hysteresis at the silica water interface.[55-56] We then computed the $\chi^{(2)}$ and $\Phi(0)$ point estimates from eqn. 8 using $\chi_X^{(3)} = 1.5 \times \chi_{water}^{(3)}$ (Fig. 4b). Argand diagrams depicting the relevant elements of eqn. 8 in the complex plane are shown in Fig. 4c, for $\chi_X^{(3)} = 1.5 \times \chi_X^{(3)}$, plotted between 10 μM and 0.5 M. Fig. 4d shows that the HD-SHG derived surface potentials are somewhat larger in magnitude than the ones calculated from Gouy-Chapman theory[5-8] for surface charge densities of -0.015 C m⁻², -0.02 C m⁻², and -0.04 C m⁻², typical for circumneutral pH. This difference may reflect the fact that the Gouy-Chapman model considers contributions to the electrostatic potential that are of Coulombic form, while dipole potentials and other contributions to the electrostatic potential are neglected. Recent experimental and computational work shows that such contributions can be considerable.[53, 57-58]

Fig. 4e shows the second-order susceptibilities for MgSO₄ are somewhat smaller in magnitude than the ones we obtain for NaCl, whereas the interfacial potentials are comparable (the measured SHG amplitudes and phases are shown in Supporting Information Fig. S7, while the Debye length for the MgSO₄ case was calculated according to ref. 38 in Majumder et al.[59]). The differences in $\chi_{MgSO_4}^{(2)}$ vs $\chi_{NaCl}^{(2)}$ are consistent with the differences in hyperpolarizabilities computed for related systems (M$^{n+}$···NH₃ complexes and M$^{n+}$ coordinated within organic macrocycles).[60-61] The results point towards the utility of HD-SHG to provide structural information on otherwise difficult-to-study ions at interfaces. This apparent ion sensitivity exhibited in Fig. 4e sets up the HD-SHG as an experimental source of structural information that can be directly compared to second-order nonlinear susceptibilities derived from atomistic simulations.[17, 20, 62-65]

Homodyne-detected vibrational sum frequency generation (SFG) spectra collected along with homodyne-detected nonresonant SHG signals were reported recently by Rehl at al.[27] at the pH and ionic strength conditions employed here. While surface potentials and second-order nonlinear susceptibilities were not reported, differences in the SFG and SHG responses were found to be substantial and attributed to interference of the silica and net-aligned water molecules. We therefore compared the interference fringe from a bare silica hemisphere (no water present) with that of the same hemisphere after filling the aqueous flow cell with deionized water at pH 5.8 (2 μM ionic strength). The resulting 180° phase shift (please see Supporting Information Fig. S8) experimentally confirms the expectation from the literature.[25, 66]

In conclusion, the experimental evidence and analysis presented here posits the existence of a third-order nonlinear susceptibility, $\chi_X^{(3)}$, associated with a 90° phase shift, that contributes to the SHG response from charged interfaces. When included in a model for the total second-order nonlinear susceptibility, reasonable point estimates for $\Phi(0)$ are obtained for $\chi_X^{(3)} \approx 1.5 \times \chi_{water}^{(3)}$. The model also produces reasonable point estimates for $\chi^{(2)}$ of the fused silica:water interface over a commonly studied range of pH values and ionic strengths. Moreover, the model has a demonstrated utility for distinguishing apparent differences in the second-order nonlinear susceptibility when the electrolyte is NaCl *vs.* MgSO₄, pointing to the possibility of using HD-SHG to investigate ion-specificity in interfacial processes.

The results highlight the need for more research, from theory and experiment, to elucidate the precise origin and magnitude of the newly identified term. Improvements to the precision and accuracy of the SHG phase measurement should also be made. In addition to its geochemical relevance, the results presented here may be of interest for nonlinear optical studies of electrochemical, energy-related, and biological interfaces through spectroscopy and microscopy.[67-72] Beyond non-resonant conditions, an implication of our findings for the interpretation of resonantly enhanced vibrational SFG spectra of charged interfaces is that even at high ionic strength, absorptive-dispersive mixing is expected to render substantial influence on the detected spectral lineshapes, as we had eluded to earlier.[16-17, 20] Applying eqn. 8 to those SFG spectra, and realization of a second heterodyne detection step in HD-SFG, hold the promise of revealing new chemistry and physics for charged interfaces.

**Methods.**

**Sample Preparation.** Fused silica hemispheres (Hyperion Optics Corning 7979 IR Grade) were used as received. First, they were soaked in Alnochromix cleaning solution (Alconox Labs) for one hour and rinsed with copious amounts of ultrapure water (Millipore Sigma, 18.2 M$\Omega \times$ cm). The cleaned hemispheres were then sonicated for 15 minutes inside a small beaker filled with methanol. Then the methanol was poured out without removing the hemisphere, and the beaker and the hemisphere were triple-rinsed with ultrapure water. The beaker was then filled with ultrapure water and the hemisphere was sonicated for another 15 minutes. After removing the hemisphere from the beaker, it was dried in house $N_2$ gas. Then, the hemisphere was plasma cleaned for 30 seconds (Harrick, "high" setting), before assembly in a previously described PTFE/aluminum sample cell[1-2, 37] using a Viton O-ring that had been previously stored in ultrapure water and not dried - *n.b.*: O-rings used for the dry experiments (fused silica:air and fused silica:$\alpha$-quartz) were rinsed in methanol, then in ultrapure water, then dried in house $N_2$ and then plasma cleaned. Prior to assembly, the PTFE component of the cell was sonicated in methanol for 15 minutes, rinsed in ultrapure water, dried under house $N_2$, and then plasma cleaned for 2 minutes. Aqueous salt solutions of 100 mM and 500 mM concentrations were prepared 24 hours in advance of the experimental measurements, and allowed to reach equilibrium with laboratory air, *i.e.* ambient $CO_2$, while being stirred with a slightly opened lid, for a final reading of pH 5.7$\pm$0.1 and a conductivity of 10 mS cm$^{-1}$ and 45 mS cm$^{-1}$, respectively, prior to use. Deionized water from the Millipore system was also stored overnight and allowed to equilibrate with laboratory air the same way as the ionic solutions for a final reading of pH. 5.7±0.1 and a conductivity of 0.6±0.2 μS cm⁻¹ prior to use. All other solutions were prepared and pH-balanced using 0.1 M stock solutions of HCl and NaOH (both from Fisher Scientific) prior to the start of each experiment.

**HD-SHG Data Collection and Analysis.** Details of the HD-SHG spectrometer have been previously described.[1-2, 37] As pointed out in our previous work, we assemble the sample cell and fused silica hemisphere and the aqueous flow lines, close the lid on the entire spectrometer/sample cell assembly, and wait for three hours. This procedure minimizes subsequent SHG phase drift. Prior to starting an experiment, the SHG intensity was monitored upon introduction of equilibrated pH 5.7 ultrapure water to the flow cell until the signal remained stable for 5 minutes. The same benchmark (signal remains stable for 5 minutes) was applied for all changes in the aqueous solutions (pH, ionic strength) that were made during a given experiment. After a stable baseline SHG signal was obtained, 5 interference fringes were collected in succession by translating a piece of $\alpha$-quartz to 30 equidistant points along an automated 100 mm translational stage. The heterodyned SHG response (LO+S) was averaged for 5 seconds at each position. This procedure takes four minutes for each scan, which includes 30 sec to 40 sec to reset the stage to the 0 mm starting position. All SHG experiments were performed under a continuous aqueous flow rate of 5 mL min⁻¹. For our sample cell geometry, this flow rate establishes shear rates <10. Switching of the various aqueous solutions occurs *via* the use of a computer controlled four-channel peristaltic pump so that the lid enclosing the spectrometer/sample cell assembly does not have to be opened during our experiments. As described earlier, the interference patterns were fit to the following expression:

$$y = y_0 + A \times cos(fx + \varphi_{fit}) \tag{11},$$

where the SHG response ($E_{sig}$) corresponds to the amplitude (A) and the experimentally measured phase ($\varphi_{sig}$) is obtained by referencing $\varphi_{fit}$ to the phase obtained at pH 2.5 and 0.5 M [NaCl]. The amplitude, in turn, is normalized to that obtained at pH 5.7 and 0.5 M [NaCl], as indicated on the y-axes of Fig. 1e and Fig. 4a. The homodyne measurements needed for the C/R ratio were carried out over seven replicates each for the fused silica:water and fused silica:α-quartz, with a brief removal of the hemisphere from the sample stage in between each replicate measurement. Following slight adjustments of the steering mirror and focal lens position in front of the sample to maximize the SHG intensity from the fused silica:water interface in between each replicate to compensate for slight changes in the position of the hemisphere, a standard error of 5 percent in the SHG intensity was obtained over the seven replicate measurements. Omitting this readjustment step led to a 20 percent standard error. The optics were not readjusted in between replicate measurements for the case of the fused silica:α-quartz interface so as to avoid walking the alignment off into the bulk :α-quartz.

**V. Associated Content**

Supporting Information: optical fringe data, finite element calculations, values of $E_{sig}$ and $\varphi_{sig}$, referencing, and derivation of $\chi^{(3)}$ contribution for a linearly decaying field.

**VII. Acknowledgement.** P.E.O. gratefully acknowledges support from the NSF GRFP, the Schmidt Science Fellowship in partnership with the Rhodes Trust, and the Harvard University Center for the Environment. H.F.W. gratefully acknowledges support from the National Natural Science Foundation of China (NSFC Grant No. 21727802). T. F. M. and F.M.G. gratefully acknowledge support from the California Institute of Technology and Northwestern University, respectively.

**Figure Captions.**

**Figure 1.** a) SHG Intensity recorded as a function of local oscillator translational stage position for fused silica surfaces in contact with water held at pH 1 and pH 5.8 at constant ionic strength of 0.5 M [NaCl] (circles) and corresponding sine fits (lines). Each mm distance corresponds to 3.158°. Each scan acquired in four minutes. b) SHG intensity recorded a as a function of input power for local oscillator (LO) + signal (S), (top, purple filled circles), LO only, (middle, grey filled circles), and signal only (bottom, filled blue circles), and fits to a function of the form $y = a \cdot x^p$. Power p and standard error from the fit are indicated. c) SHG Intensity recorded as a function of local oscillator translational stage position for fused silica surfaces for pH 1-12, and sign fits using eqn. 11. d) SHG Phase and amplitude obtained from the sine fits as a function of bulk solution pH at constant 0.5 M [NaCl]. e) Replicate measurements of the SHG phase and amplitude for pH 5.8, as indicated in the dashed box of Fig. 1d), with shaded area indicating standard error of the mean. The SHG phase is referenced to pH 2.5 at 0.5 M NaCl, and the amplitude is normalized to pH 5.8 at 0.5 M NaCl. f) Newly identified nonlinear susceptibility plotted against surface potentials determined from impedance measurements (circles) and linear least-squared fit (solid line). Please see text for details.

**Figure 2.** a, b) Electrostatic field across the fused silica:water interface from finite element calculations. The silica contains 100 ppm of charged impurities and the aqueous phase is at 0.1 M [NaCl]. The relative permittivity in the aqueous phase is 78. c) Height vs position profiles from several atomic force microscopy lines scans across the flat side of our fused silica hemispheres, primitive cell used for finite element calculations shown in black box nested in between its periodic images, and electrostatic field across an atomically smooth (black line) and rough (purple line for L= H=1 nm features, blue line for L=H=2 nm features) fused silica:water interface. Horizontal

dashed line depicts the E-field value used for the 1-nm wide rough region. The relative permittivity in the first nm of the aqueous phase is 3, while it is 78 elsewhere in the water. The surface charge density is -0.015 C m$^{-2}$.

**Figure 3.** a) Point estimates of $\chi^{(2)}$ and $\Phi(0)$ obtained from eqn. 10 as a function of bulk solution pH at a constant ionic strength of 0.5 M. Shaded areas indicate 5% uncertainties. Results are for $\chi_X^{(3)} = \chi_X^{(3)} = 1.5 \times \chi_{water}^{(3)}$. Dashed line (pH1-11) and solid line (pH 1-5) are from published impedance measurements performed at comparable ionic strength. Please see text for details. b) Argand diagrams of the measured SHG amplitude ($E_{sig}$) and SHG phase angle ($\varphi_{sig}$) and the point estimates of $\chi^{(2)}$ and the $\chi_{water}^{(3)} \times \Phi(0) \times \cos(\varphi_{DC})$ and $\Phi(0) \times \chi_X^{(3)}$ products for $\chi_X^{(3)} = 1.5 \times \chi_{water}^{(3)}$. Magnitudes for pH 1 are multiplied by ten in inset.

**Figure 4.** a) SHG Amplitude and phase recorded at fused silica:water interfaces subjected to different bulk ionic strengths at pH 5.8=constant. The SHG phase is referenced to pH 2.5 at 0.5 M NaCl, and the amplitude is normalized to pH 5.8 at 0.5 M NaCl. b) Point estimates of $\chi^{(2)}$ and $\Phi(0)$ obtained from eqn. 10 as a function of bulk ionic strength at constant pH 5.8. Shaded areas indicate 10% uncertainties. Dashed line indicates Gouy-Chapman result using a charge density of -0.015 C m$^{-2}$. c) Argand diagrams of the measured SHG amplitude ($E_{sig}$) and SHG phase angle ($\varphi_{sig}$) and the point estimates of $\chi^{(2)}$ and the $\chi_{water}^{(3)} \times \Phi(0) \times \cos(\varphi_{DC})$ and $\Phi(0) \times \chi_X^{(3)}$ products for $\chi_X^{(3)} = 1.5 \times \chi_{water}^{(3)}$. d) Gouy-Chapman potential calculated for ionic strengths indicated using a surface charge density of -0.015 C m$^{-2}$ (green circles), -0.02 C m$^{-2}$ (light blue circles), and -0.04 C m$^{-2}$ (dark blue circles) plotted against the point estimates of $\Phi(0)$ obtained by HD-SHG at the same ionic strengths. Dashed line indicates 1:1 correlation. e) Point estimates of $\chi^{(2)}$ and $\Phi(0)$

obtained from eqn. 10 as a function of bulk ionic strength of $MgSO_4$ at constant pH 5.8 for $\chi_X^{(3)}$

=1.5× $\chi_{water}^{(3)}$. Shaded areas indicate 5% uncertainties.

**Figure 1**

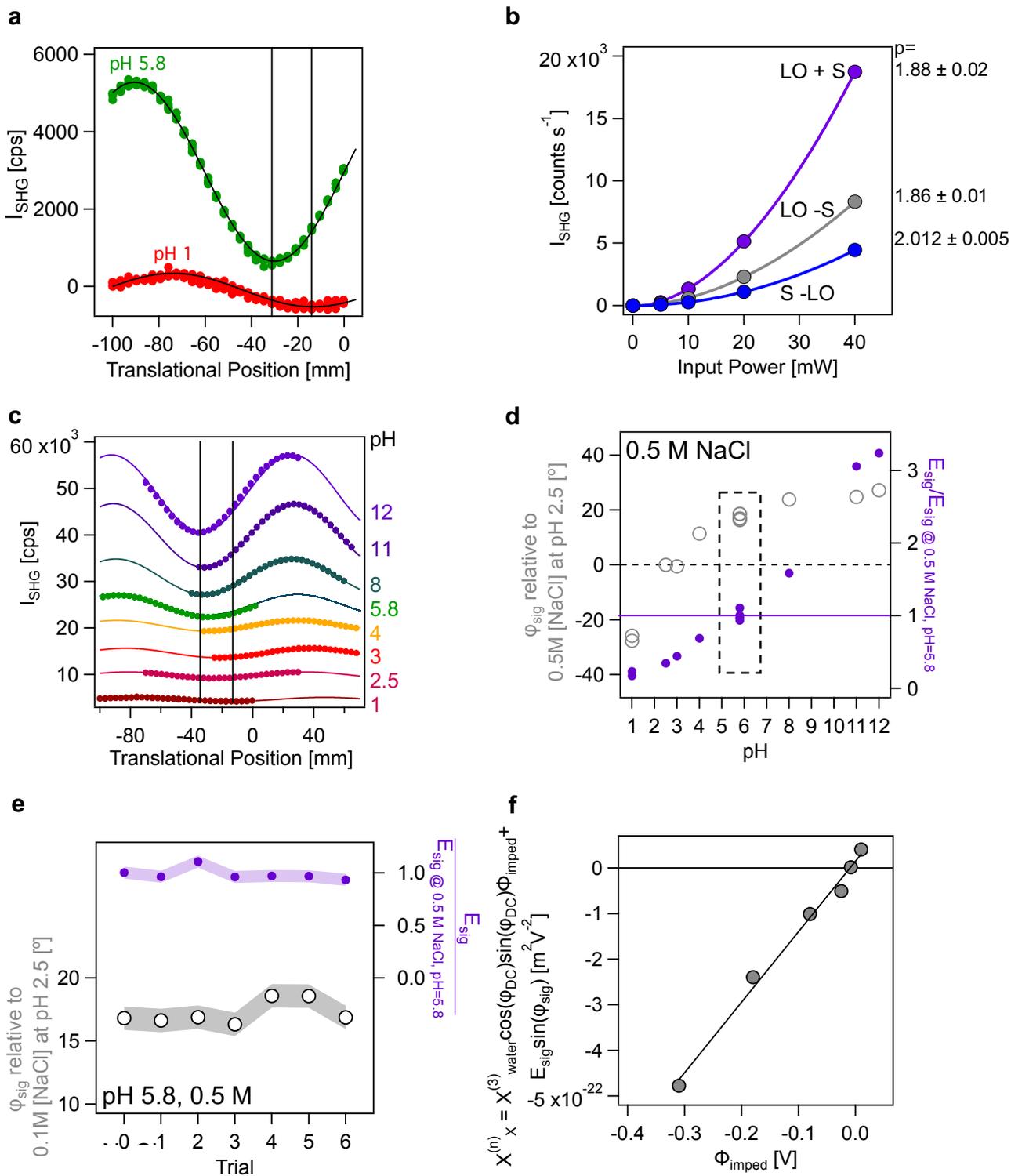

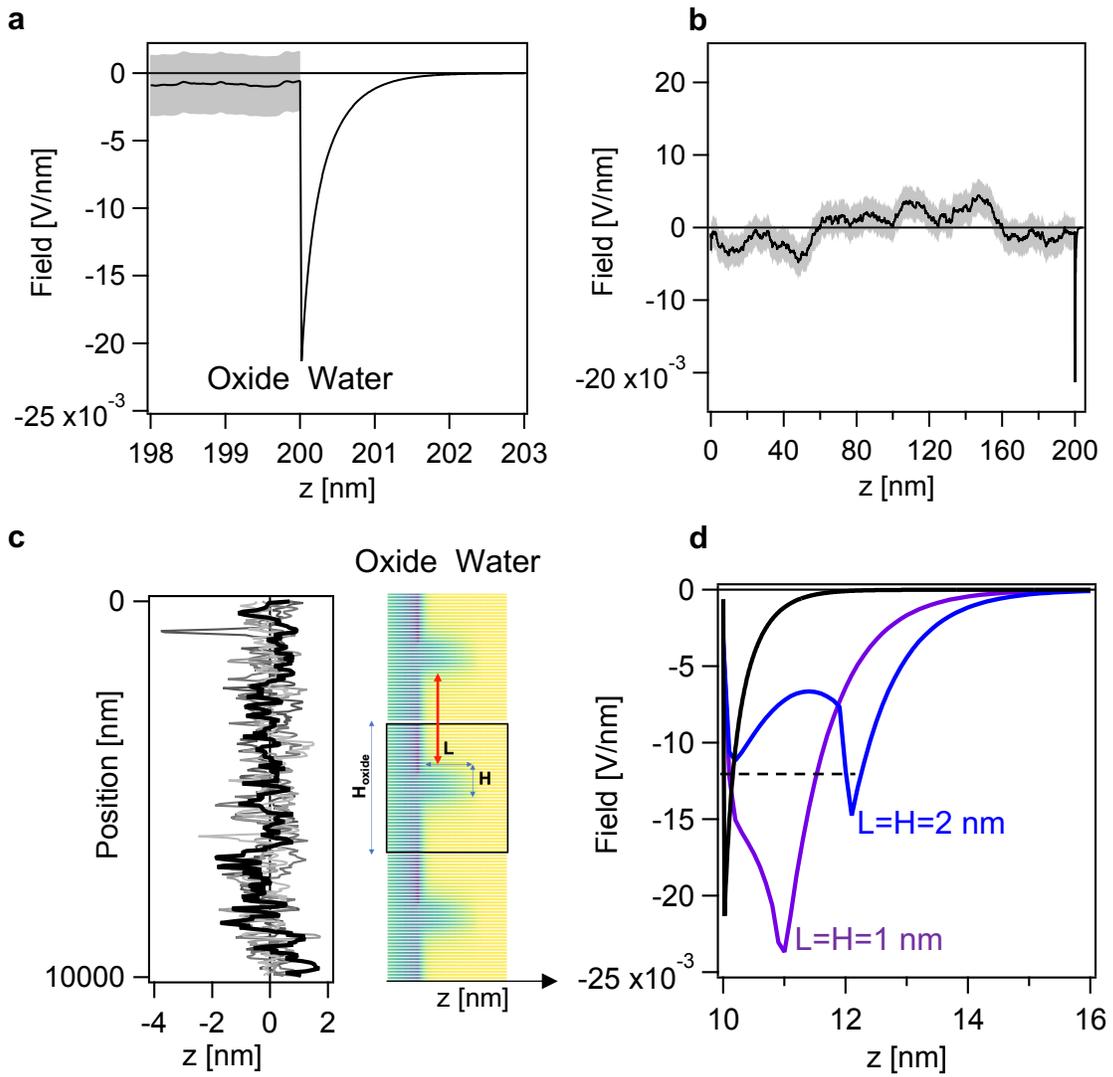

**a**

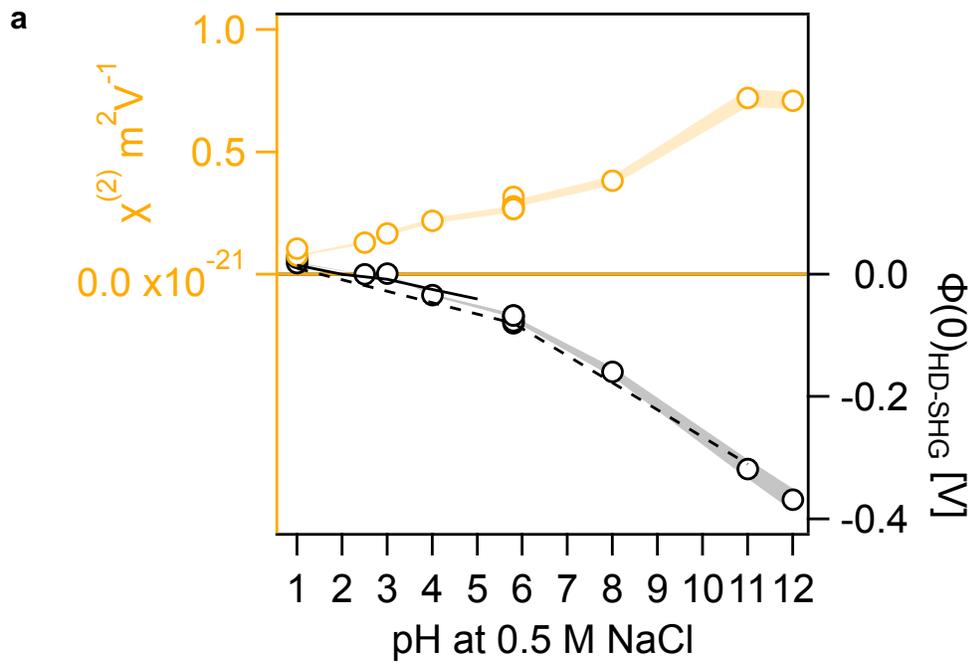

**b**

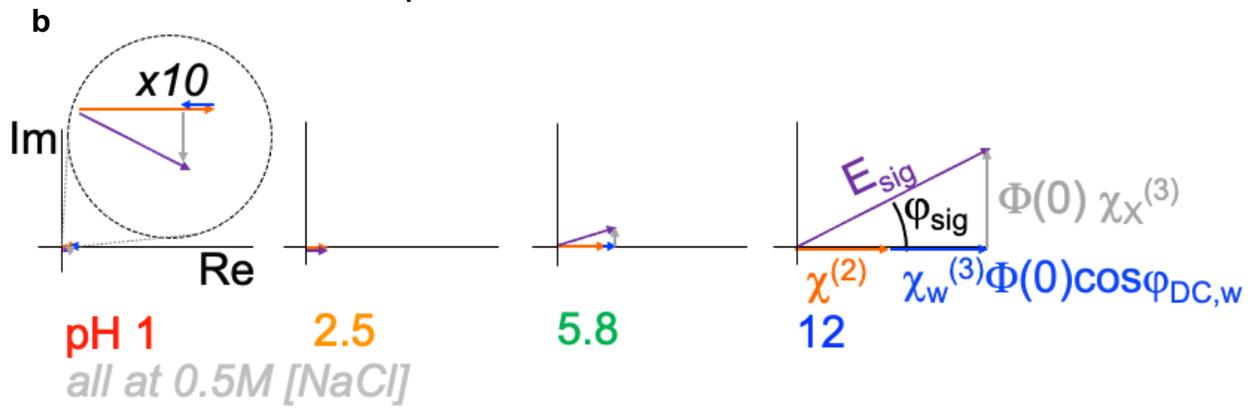

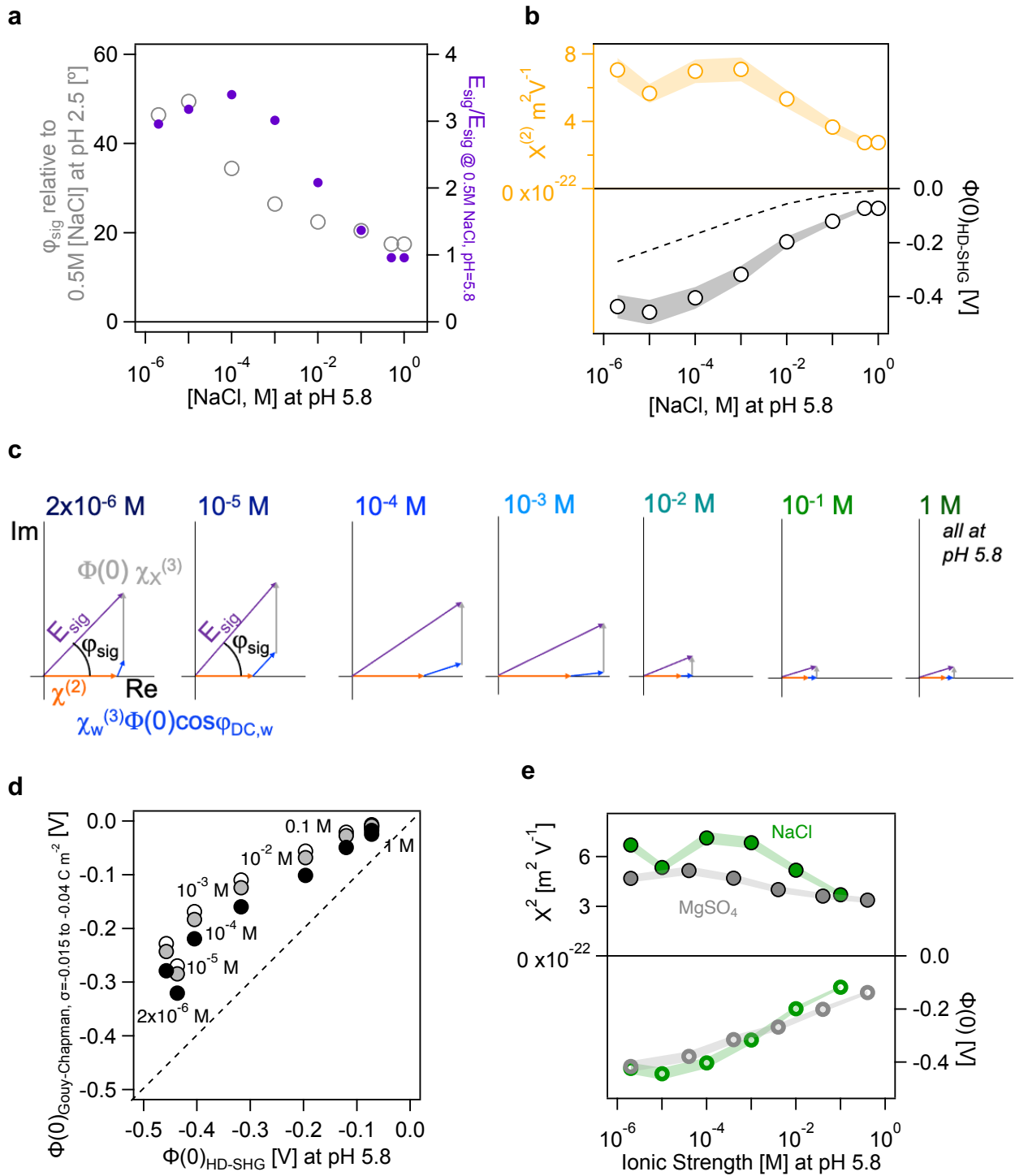

**TOC Graphic**

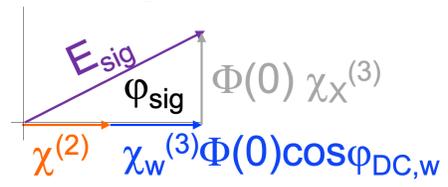

Supporting Information for

**A New Imaginary Term in the 2ⁿᵈ Order Nonlinear Susceptibility from Charged Interfaces**

Emily Ma,[#] Paul E. Ohno,[$] Jeongmin Kim,[%] Yangdongling Dawning Liu,[#] Emilie H. Lozier,[#]

Thomas F. Miller III,[%] Hong-Fei Wang,[†@] and Franz M. Geiger[#]*

[#]Department of Chemistry, Northwestern University, 2145 Sheridan Road, Evanston, IL 60660,

USA

[$]Harvard University Center of the Environment, Harvard University, Cambridge, MA 02138,

USA

[%]Division of Chemistry and Chemical Engineering, California Institute of Technology,

Pasadena, CA 91125, USA

[†]Department of Chemistry and Shanghai Key Laboratory of Molecular Catalysis and Innovative

Materials, Fudan University, 220 Handan Road, Shanghai 200433, China

[@]School of Sciences, Westlake University, Shilongshan Road No.18, Cloud Town, Xihu District,

Hangzhou, Zhejiang 310024, China

*Corresponding author: geigerf@chem.northwestern.edu

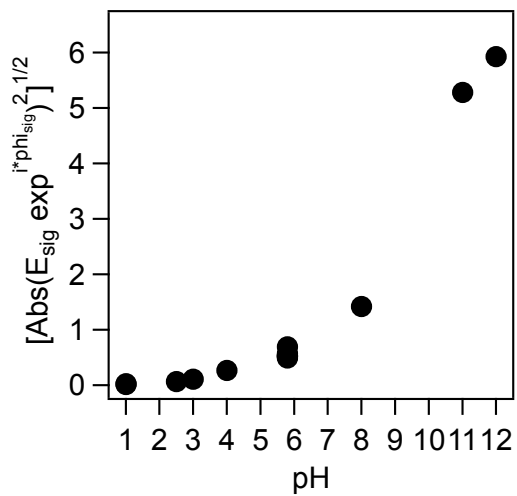

**Fig. S1**. Square root of the square modulus of the SHG amplitude and phase.

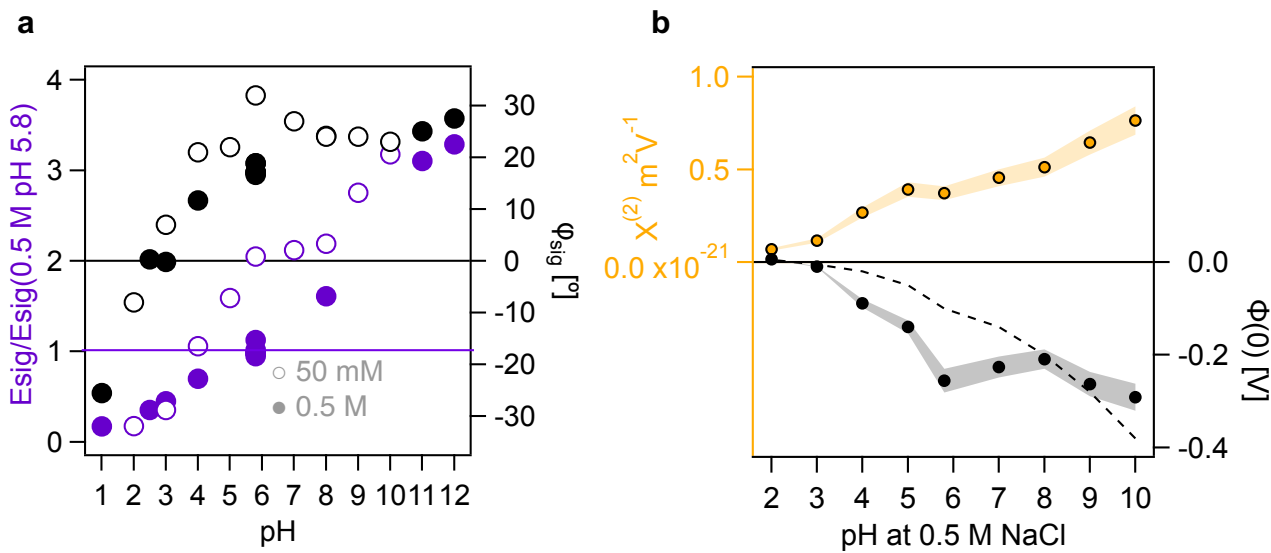

**Fig. S2.** a) SHG Phase and amplitude obtained from the sine fits as a function of bulk solution pH at constant 0.5 M NaCl (filled black and purple circles) and constant 50 mM NaCl (empty black and purple circles). The SHG phases are referenced to the one obtained at pH 2.5 at 0.1 M NaCl. The amplitudes are normalized to the one obtained at 0.5 M at pH 5.8. b) Point estimates of $\chi^{(2)}$ and $\Phi(0)$ obtained from eqn. 10 as a function of pH at 50 mM [NaCl]. Shaded areas indicate 10% uncertainties. Dashed lines are the X-ray spectroscopic surface potentials for silica colloids suspended in a liquid jet at 50 mM [NaCl] reported by Brown et al.[1]

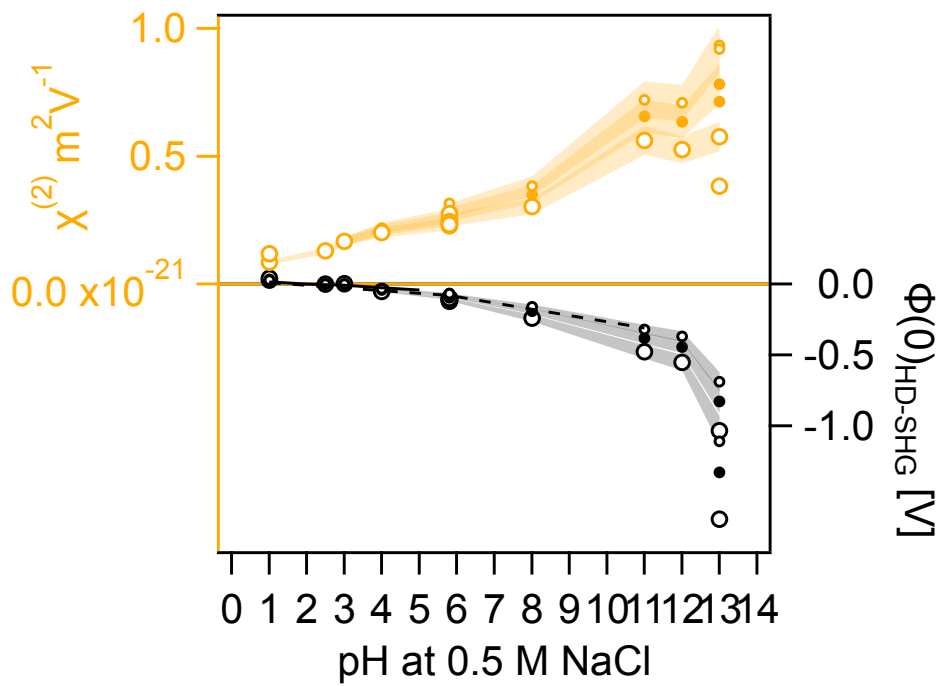

**Fig. S3.** Point estimates of $\chi^{(2)}$ and $\Phi(0)$ obtained from eqn. 10 as a function of bulk ionic strength at constant pH 5.8. Shaded areas indicate 10% uncertainties, for full pH range studied.

**Supporting Information Note S1.** The 3$^{\text{rd}}$-order silica contribution in eqn. 4 is derived *via*

$$\chi_{dc}^{(2)} = \int_{-\infty}^{0} \chi^{(3)} E_{DC}(z) e^{i\Delta k_z z} dz = \int_{-\infty}^{0} -\chi^{(3)} \frac{d\Phi(z)}{dz} e^{i\Delta k_z z} dz = \int_{-\infty}^{0} -\chi^{(3)} e^{i\Delta k_z z} d\Phi(z) \quad \text{(S1)},$$

where we apply a linearly decaying potential into the bulk silica according to $\Phi(z) = a(z + b)$ $(for -b \le z \le 0)$ and $\Phi(z) = 0$ $(for z \le -b)$. We then obtain

$$\chi_{dc}^{(2)} = \int_{-b}^{0} -\chi^{(3)} e^{i\Delta k_z z} a\, dz = \int_{e^{-i\Delta k_z b}}^{1} -\frac{\chi^{(3)} a}{i\Delta k_z} d e^{i\Delta k_z z} = \frac{\chi^{(3)} a}{i\Delta k_z} \left( e^{-i\Delta k_z b} - 1 \right) \tag{S2}$$

$$= \frac{a\chi^{(3)}}{i\Delta k_z} (\cos \Delta k_z b - 1 - i \sin \Delta k_z b) \tag{S3}$$

$$= -\frac{a}{\Delta k_z} [(\sin \Delta k_z b) - i(1 - \cos \Delta k_z b)] \chi^{(3)} \tag{S4}$$

$$= -\frac{a}{\Delta k_z} \sqrt{(\sin \Delta k_z b)^2 + (1 - \cos \Delta k_z b)^2} \chi^{(3)} e^{-i\varphi_{DC}} \tag{S5}$$

$$= -\frac{a}{\Delta k_z} \sqrt{(\sin \Delta k_z b)^2 + (\cos \Delta k_z b)^2 - 2 \cos \Delta k_z b + 1} \chi^{(3)} e^{-i\varphi_{DC}}$$

$$= -\frac{a}{\Delta k_z} \sqrt{2 \times (1 - \cos \Delta k_z b)} \chi^{(3)} e^{-i\varphi_{DC}} \tag{S6}$$

$$= -\frac{a}{\Delta k_z} \sqrt{2 \times 2 \left( \sin \frac{\Delta k_z b}{2} \right)^2} \chi^{(3)} e^{-i\varphi_{DC}} = -\frac{2a}{\Delta k_z} \left| \sin \frac{\Delta k_z b}{2} \right| \chi^{(3)} e^{-i\varphi_{DC}} \tag{S7}$$

$$= -\frac{2a}{\Delta k_z} \left( \sin \frac{\Delta k_z b}{2} \right) \chi^{(3)} e^{-i\varphi_{DC}} \tag{S8}$$

Since $\Phi(0) = ab$, $a = \frac{\Phi(0)}{b}$. We therefore obtain

$$\chi_{dc}^{(2)} = -\frac{2}{\Delta k_z} \cdot \frac{\Phi(0)}{b} \left( \sin \frac{\Delta k_z b}{2} \right) \chi^{(3)} e^{-i\varphi_{DC}} = -\frac{\Phi(0)}{\frac{\Delta k_z b}{2}} \left( \sin \frac{\Delta k_z b}{2} \right) \chi^{(3)} e^{-i\varphi_{DC}} \quad \text{(S10)},$$

where $\varphi_{DC} = \arctan \left( \frac{1 - \cos \Delta k_z b}{\sin \Delta k_z b} \right) = \arctan \left( \tan \frac{\Delta k_z b}{2} \right) = \frac{\Delta k_z b}{2}$.

Therefore, $\chi_{dc}^{(2)} = -\frac{\sin \varphi_{DC}}{\varphi_{DC}} \chi^{(3)} \Phi(0) e^{-i\varphi_{DC}}$, where $\varphi_{DC} = \Delta k_z \frac{b}{2} = \Delta k_z \lambda_D$.

**Supporting Information Note S2.** The one-dimensional Poisson equation was numerically solved

for an oxide in contact with ionic water of 0.1 mol L$^{-1}$ NaCl using the Newton-Raphson iteration

method:[2]

$$\frac{d^2\Psi(x)}{dx^2} = \begin{cases} -\frac{e[\delta(x_p)-\delta(x_n)]}{\epsilon_{ox}\epsilon_0} & \text{if } x < L_{ox} \\ -\frac{e\rho_b}{\epsilon_w\epsilon_0}\left[e^{-e\beta\Psi} + e^{e\beta\Psi}\right] & \text{if } L_{ox} < x < L \end{cases} \qquad \text{(eqn. S11)}.$$

Here, $\Psi(x)$ is the one-dimensional electrostatic potential, and $L$ (= 220 nm) is the total length of the system, including a 200 nm thick oxide and a 20 nm thick water region. The water-oxide interface is located at $x = L_{ox} = 200$ nm. The vacuum permittivity is $\epsilon_0 = 8.854 \times 10^{-12}$ Fm$^{-1}$, the relative dielectric permittivities are $\epsilon_{ox} = 4$ for the oxide and $\epsilon_w = 78$ for the ionic water, $e$

is the elementary charge, and $b=k_B T^{-1}$ with T being the temperature (300 K) and $k_B$ being the

Boltzmann constant. The bulk ion density in the ionic water is $\rho_b = 0.1$ mol L$^{-1}$ with monovalent cations and anions whose thermal motion is incorporated using the Boltzmann factor.

The oxide is modeled to include a pair of positively and negatively charged point defects

that decrease the electric field inside the oxide. A positively charged defect ($x = x_p$) is randomly

placed inside the oxide region, followed by a random insertion of a negatively charged defect ($x =$

$x_n$) according to a Poisson distribution having an average of 100 nm, so the distance between the

defects is 100 nm on average to recapitulate the 100 ppm defect density in commercially available silica.

Three boundary conditions are applied in our model using the fact that the oxide surface in

contact with water is lightly charged and both the oxide and water boundaries are electrically

grounded. We employ (i) the Dirichlet condition with $\Psi(x = 0) = 0$ at the oxide end, (ii) the

Dirichlet condition with $\Psi(x = L) = 0$ at the water end, and (iii) the Neumann condition at the

aqueous oxide interface ($x = L_{ox}$) as follows:

$$\epsilon_{ox} \frac{d\Psi(x)}{dx}\Big|_{x=Lox} - \epsilon_w \frac{d\Psi(x)}{dx}\Big|_{x=Lox} = \frac{Q_s}{\epsilon_0}, \qquad \text{(eqn. S12)}$$

where $Q_s$ (= -0.015 C m$^{-2}$) is the oxide surface charge density. The differential equation is solved by discretizing the space with finite elements whose size is 0.01 nm along the x-axis. Note that 0.01 nm is short enough to properly recapitulate the Debye screening length (0.39 nm at 1 mol L$^{-1}$ NaCl) in the water region. The iteration is terminated once L$_2$-norm of the solution is less than 10$^{-6}$. The electric potential and the associated electric field is averaged over 10$^6$ realizations with the random insertion of defects inside the oxide.

The two-dimensional linearized Poisson-Boltzmann equation was numerically solved for rough oxide:water interfaces (See Fig. S XX), modeled by a small dendrite of height (H) and length (L):

$$\left(\frac{d^2}{dx^2} + \frac{d^2}{dy^2}\right)\Psi(x,y) = \begin{cases} 0, \text{in the oxide region} \\ \kappa^2\Psi(x,y), \text{in the water region} \end{cases} \qquad \text{(eqn. S13)},$$

where $\kappa$ (= $\sqrt{\frac{2e^2\beta\rho_b}{\epsilon_w\epsilon_0}}$) is the inverse of Debye screening length and $\Psi(x,y)$ is the two-dimensional electrostatic potential. Here, the length of each region is fixed with a 10 nm thick oxide and a 20 nm thick water region. The height of the oxide region varies according to the height of the dendrite: $H_{oxide} = 2H$. Other parameters are kept the same as in the one-dimensional model above.

Boundary conditions are applied as in the one-dimensional model. We employ (i) the Dirichlet condition with $\Psi(x,y) = 0$ at both oxide and water ends (blue line in Fig. S4), and (ii) the Neumann condition at the corrugated aqueous oxide interface, $\partial$, (yellow line in Fig. S4) as follows:

$$\epsilon_{ox}\left(\frac{d}{dx} + \frac{d}{dy}\right)\Psi_{ox}(x,y)\Big|_{\partial} \cdot \hat{n} - \epsilon_w\left(\frac{d}{dx} + \frac{d}{dy}\right)\Psi_w(x,y)\Big|_{\partial} \cdot \hat{n} = \frac{Q_s}{\epsilon_0} \qquad \text{(eqn. S14)}.$$

Here, $\hat{n}$ is the unit normal vector pointing the water region from the oxide region. The differential equation is solved by discretizing the space with finite elements whose size is 0.1 nm along both x- and y-axes. Note that 0.1 nm is still short enough to properly recapitulate the Debye screening length (0.39 nm at 0.1 mol L$^{-1}$ NaCl) in the water region. The linear equation is solved using Jacobi and Gauss-Seidel methods with successive over-relaxation.[3] The relaxation method is terminated once L$_2$-norm of the solution is less than $10^{-12}$.

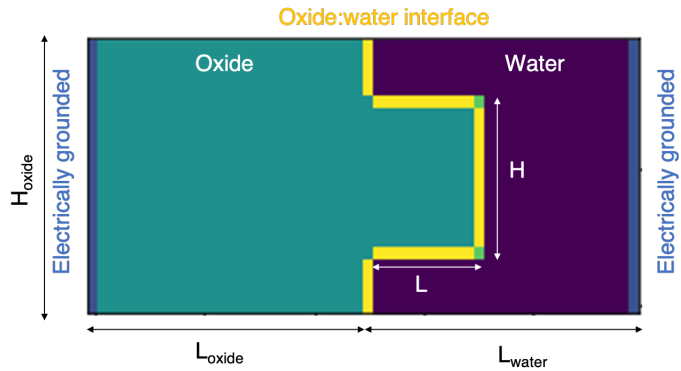

**Fig. S4.  Finite element calculation model for a corrugated oxide:water interface.** The rough oxide:water interface is modeled using a nanoscale dendrite of height (H) and length (L), represented by the yellow region. Bright green elements are the dendrite corners where the oxide:water boundary is along both the x- and y-axes.

**Supporting Information Note S3.**

We begin with

$$\chi_{tot}^{(2)} = \chi^{(2)} \pm i\chi_q^{(2)} - \Phi(0)\chi_{EDL}^{(3)} \cos\left(\varphi_{DC,EDL}\right) e^{i\varphi_{DC,EDL}} = E_{sig} \times e^{i\varphi_{sig}} \quad \text{(S15a)}$$

$$= E_{sig} \times \cos\varphi_{sig} + E_{sig} \times i \times \sin\varphi_{sig} \quad \text{(S15b)}.$$

Collection of the real and imaginary parts yields

$$E_{sig} \times \sin\varphi_{sig} = \chi_q^{(2)} - \Phi(0)\chi_{EDL}^{(3)} \cos\left(\varphi_{DC,EDL}\right) \sin\left(\varphi_{DC,EDL}\right) \quad \text{(S16) and}$$

$$E_{sig} \times cos\, \varphi_{sig} = \chi^{(2)} + \Phi(0)\chi_{EDL}^{(3)}\big(cos(\varphi_{DC,EDL})\big)^2 \qquad (S17).$$

Therefore,

$$\Phi(0) = \frac{-\big(E_{sig} \times sin\, \varphi_{sig} - \chi_q^{(2)}\big)}{\chi_{EDL}^{(3)}\, cos(\varphi_{DC,EDL})\, sin(\varphi_{DC,EDL})} \qquad (S18)\ \text{and}$$

$$\chi^{(2)} = E_{sig} \times cos\, \varphi_{sig} + \Phi(0)\chi_{EDL}^{(3)}\big(cos(\varphi_{DC,EDL})\big)^2 \qquad (S19).$$

With $\chi_q^{(2)} = 1.75 \times 10^{-22}\ m^2 V^{-1}$,[4] we find nonphysical results for $\Phi(0)$ and for $\chi^{(2)}$.

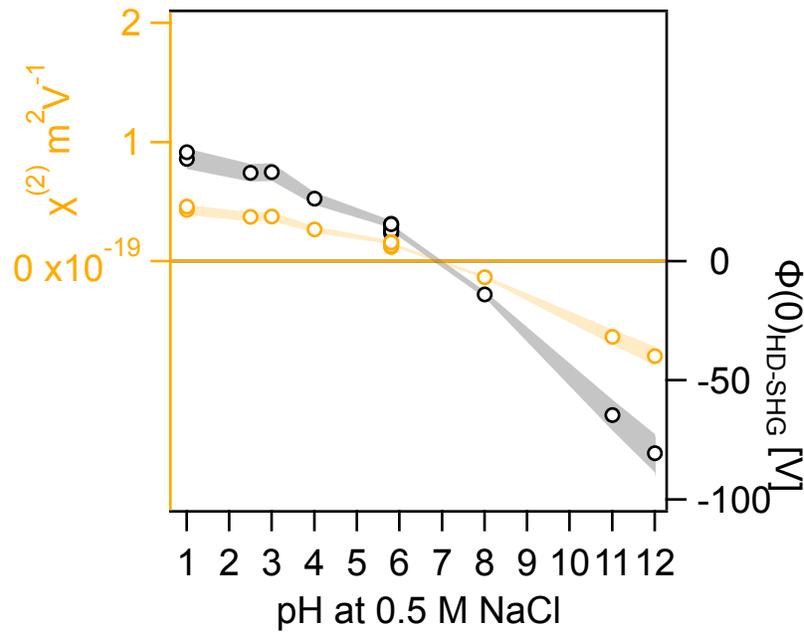

**Fig. S5.** Point estimates of $\chi^{(2)}$ and $\Phi(0)$ obtained from eqn. 8 as a function of bulk ionic strength at constant pH 5.8 for a model containing only an electrical quadrupolar $\chi_q^{(2)}$ as a new addition to model (1).

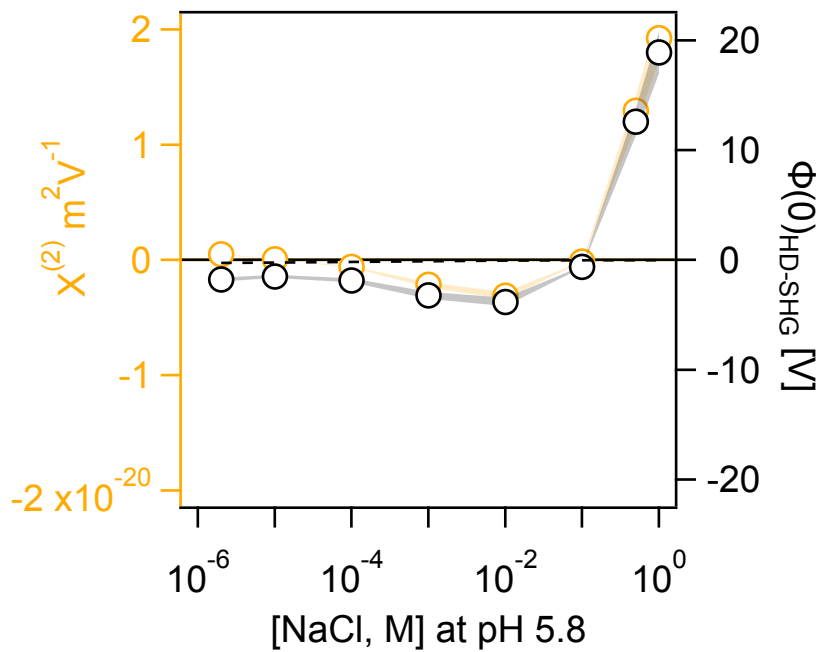

**Fig. S6.** Point estimates of $\chi^{(2)}$ and $\Phi(0)$ obtained from eqn. 8 as a function of bulk ionic strength at constant pH 5.8 for a model containing only an electrical quadrupolar $\chi_q^{(2)}$ as a new addition to model (1).

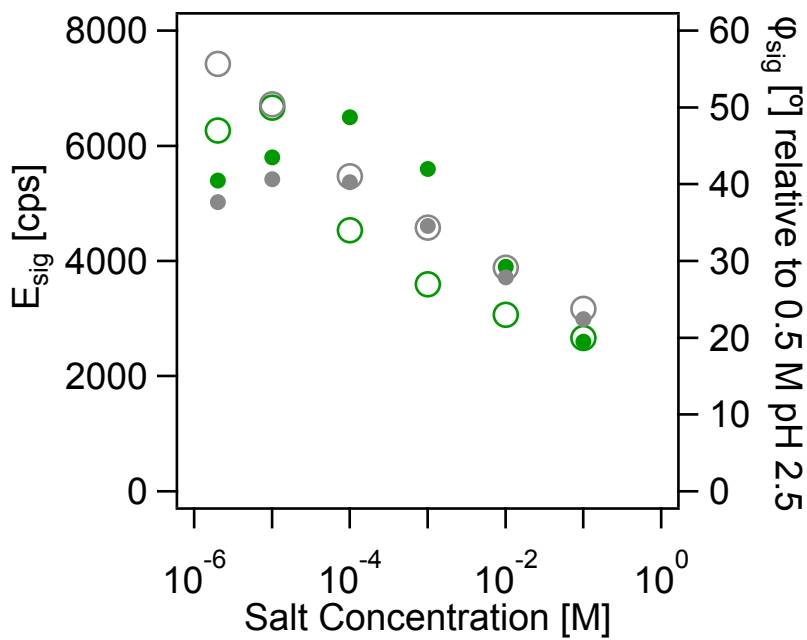

**Fig. S7.** SHG Amplitude (filed circles) and phase (empty circles) recorded from fused silica in contact with water maintained at different bulk ionic strengths of $Na_2SO_4$ (yellow) and NaCl (green) at constant pH 5.8. The SHG phase is with reference to pH 2.5 at 0.5 M NaCl .

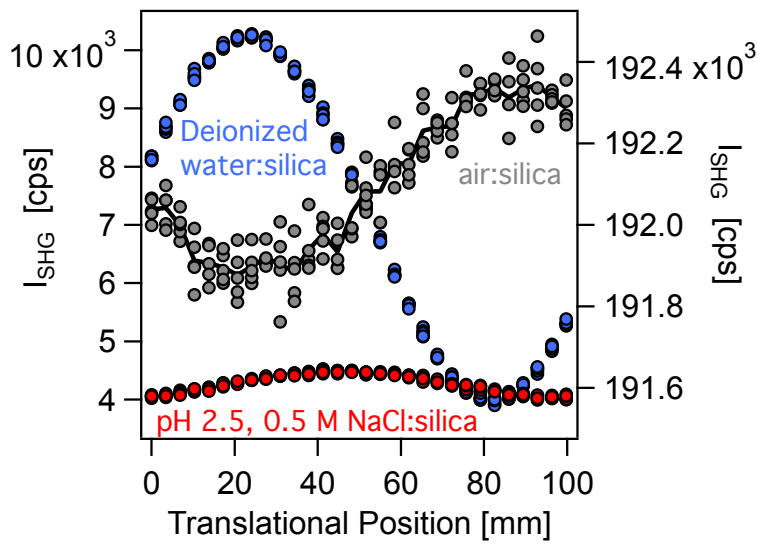

**Fig. S8.** SHG Intensity recorded as a function of local oscillator translational stage position for an air:silica interface (gray circles, and average of five fringes as black line), a water:silica interface held at pH 5.8 and 2 μM ionic strength (blue circles), and at pH 2.5 and 500 mM ionic strength (red circles). Please see text for details.

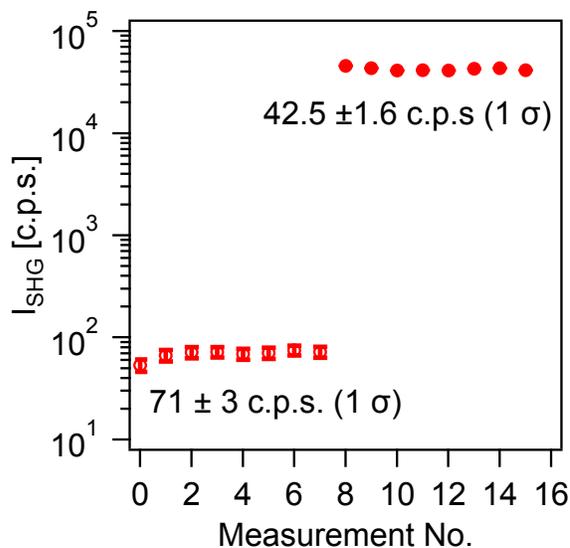

**Fig. S9.** SHG Intensity recorded for seven replicate measurements from individually assembled silica:ultrapure water interfaces (air-equilibrated overnight, 20 μS cm⁻¹ conductivity, pH 5.8) and seven subsequent replicates from individually assembled silica:α-quartz interfaces (5 mW laser power at the sample).